\documentclass[aps,prl,reprint]{revtex4-2}
\usepackage{amsmath,amssymb}
\usepackage{graphicx}
\usepackage{siunitx}
\usepackage[colorlinks=true, allcolors=blue]{hyperref}
\renewcommand

\DeclareMathOperator{\dif}{d\!}

\begin{document}

\title{Wakefield regeneration in a resonantly driven plasma accelerator}
\author{J. P. Farmer}
\email{j.farmer@cern.ch}
\affiliation{Max-Planck-Institut für Physik, München, Germany}
\author{G. Zevi Della Porta}
\affiliation{CERN, Gen\`eve, Switzerland}
\affiliation{Max-Planck-Institut für Physik, München, Germany}

\begin{abstract}
Plasma wakefields offer high acceleration gradients, orders of magnitude larger than conventional RF accelerators. However, the achievable luminosity remains relatively low, typically limited by repetition rate and the charge accelerated per shot. In this work, we show that a train of drive bunches can be harnessed to accelerate multiple witness bunches in a single shot.  We demonstrate that periodically loading the wakefields removes the limit on the energy transfer from the drive beam to the plasma, which allows the luminosity to be increased.  Proof-of-concept simulations for the AWAKE scheme are carried out to demonstrate the technique, achieving a doubling of the accelerated charge while exploiting only a fraction of the drive train.
\end{abstract}

\maketitle


Plasma-wakefield acceleration offers accelerating gradients orders of magnitude larger than conventional RF accelerating systems. Typically, a short driver, either a laser pulse or charged particle bunch, is used to excite a plasma wave. A trailing witness bunch with the correct delay will be accelerated \cite{lwfa-tajimadawson,pwfa-chen}. Over recent years, huge progress has been made in terms of the attainable beam quality \cite{pwfa-lindstrom-energyspread}. However, the average accelerated current in wakefield accelerators is typically rather low~\cite{pwfa-dArcy-recovery}.  Significant development is therefore necessary for wakefield accelerators to be competitive with RF systems in the key area of luminosity.

Short drivers (with length $L_D\sim 1/k_p=c/\omega_p$, with $1/k_p$ the plasma skin depth, $c$ the speed of light and $\omega_p$ the plasma frequency) are typically preferred due to the relative simplicity of the scheme.  
However, several schemes using a train of bunches have been investigated.  Periodic drivers were first considered for laser-driven wakefields before the availability of femtosecond lasers, achieved either through a laser-beat~\cite{lwfa-tajimadawson} or through the self-modulation of a long laser pulse~\cite{lwfa-andreev-sm,lwfa-krall-sm}.
Experiments have been carried out to demonstrate that trains of particle bunches~\cite{pwfa-muggli-multibunch} or laser pulses~\cite{lwfa_cowley_multipulse} with a periodicity $\sim2\pi/k_p$ act to resonantly excite plasma wakefields, allowing for the acceleration of a witness bunch.  The AWAKE experiment at CERN has demonstrated a proton-driven scheme, again exploiting self-modulation due to the lack of a suitably short driver~\cite{pwfa-caldwell-selfmodulation,pwfa-AWAKE-modulation}.  However, the accelerating gradient in these resonantly driven schemes is limited, as plasma nonlinearities at high wakefield amplitude break the resonance between the wake and the beam~\cite{plasma-rosenbluth-saturation,pwfa-lotov-twodimensionalequidistant}.  For pre-modulated beams, this can be avoided through careful shaping of the drive train~\cite{lwfa-umstadter-optimisedlasertrain,pwfa-chiadroni-comb,lwfa-wettering-modulated}, and in the case of a self-modulated beam, the loss of resonance can be partially compensated by using a plasma density step~\cite{pwfa-lotov-step}.

Using a train of drive bunches offers several possible advantages.  Periodic laser drivers have recently become the subject of renewed focus due to the higher efficiency with which such beams can be generated~\cite{lwfa-jakobsson-modulated}.  Using a ramped bunch train to drive the wakefields has been suggested as a method to increase the transformer ratio~\cite{pwfa-schutt-rampedtrain,pwfa-power-ramped}, allowing the per-particle witness energy to exceed that of the driver.  The creation of such beams has also been demonstrated in experiment~\cite{pwfa-chiadroni-comb}.

A train of witness bunches also has potential benefits.  For moderately nonlinear wakefields, the response of the plasma electrons to the witness bunch can result in nonlinear focussing forces, leading to emittance growth.  Spreading the witness bunch over several wakefield periods reduces the charge density, and so the nonlinearity~\cite{pwfa-katsouleas-beamloading}.  In modern schemes for electron acceleration, this is typically not necessary as acceleration takes place in a plasma blowout or ``bubble'', a cavitated wake free of plasma electrons~\cite{pwfa-rosenzweig-blowout,lwfa-pukhov-broken}.  Even when the driver does not generate a blowout, an electron witness bunch can drive its own~\cite{pwfa-olsen-inject}.  However, a train of witness bunches may still offer benefits when the response of the plasma to the witness remains important, for example the response of the plasma electrons to a positron bunch~\cite{pwfa-hue-positronequilibrium}, or for very high density electron bunches where the motion of plasma ions may become significant~\cite{pwfa-rosenzweig-ionmotion}.  Using a train of witness bunches has also been suggested as a mechanism to allow the acceleration of short bunches while maintaining the overall efficiency~\cite{pwfa-shroeder-nearhollow}.

A periodic train of interleaved drive and witness bunches has also been suggested as a method to increase the charge which can be accelerated ``per shot'', replenishing the depleted wake after each witness bunch~\cite{pwfa-meer-beamloading}.  In this paper, we revisit this concept and discuss a key benefit of this technique which has not previously been considered: periodically loading and replenishing the wakefields avoids the plasma nonlinearities which lead to saturation.  Controlling the wakefield amplitude in a resonantly driven wakefield accelerator through beamloading therefore removes the constraint on the total power transfer from a driver to the plasma.  Furthermore, the scheme can be applied to a nonuniform drive train by correctly tailoring the witness train.  Proof-of-concept simulations based on the AWAKE experiment at CERN are used to show that the injection of multiple witness bunches allows the accelerated charge per drive train to be significantly increased.  The potential application and impact of this technique is then discussed.

\begin{figure}
    \centering
    \includegraphics[width=\linewidth]{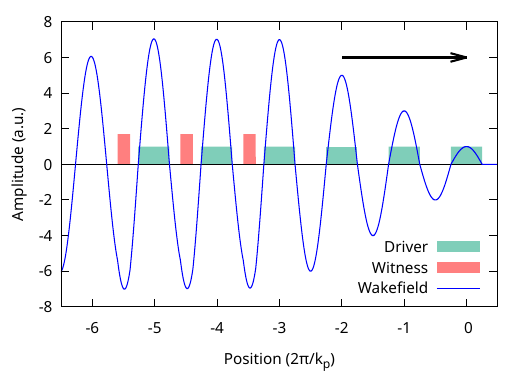}
    \caption{A train of drive bunches (green), propagating to the right, excites a plasma wakefield (blue).  By correctly loading the wakefields, a train of witness bunches can be accelerated to the same energy.}
    \label{fig:schematic}
\end{figure}

A simple schematic of the generalized scheme is shown in Fig.~\ref{fig:schematic}, with a train of drive bunches and the linear plasma response superimposed.  The limitations of linear theory will be discussed below. The accelerating field behind the first bunch is larger than the decelerating field acting on the bunch, allowing the per-particle witness energy to exceed that of the driver.  The ratio between the accelerating and decelerating fields is known as the transformer ratio, and can reach up to two for a symmetric driver~\cite{pwfa-ruth-transformer}.  Each drive bunch excites a wakefield, with the bunches separated by $2\pi/k_p$ such that the wakes sum coherently.  This configuration leads to high accelerating fields, which scale with the number of drive bunches.  However, the decelerating field acting on each drive bunch also increases along the drive train, so the transformer ratio is asymptotic to unity.  In schemes where acceleration is limited by depletion of the driver, this results in a lower efficiency~\cite{pwfa-madea-train}.  Despite this, the periodic driver still offers advantages where short drivers do not exist, or where long drive trains can be generated more efficiently~\cite{lwfa-jakobsson-modulated}.  

A witness bunch may then be injected into the wakefield.  In order for it to be accelerated, the wake it generates should destructively interfere with that excited by the driver.  The plasma wakefield is then suppressed while the witness gains energy.  This loading of the wakefields can be seen behind the first microbunch in the witness train in Fig.~\ref{fig:schematic}.

If the witness bunch is injected within the drive train, the following drive bunch will be in phase with the depleted wakefield, allowing the wakefield to be regenerated.  In this way, a train of witness bunches can be accelerated.  In order to increase the luminosity of the accelerator, each witness bunch must be accelerated to the same energy.  This can be achieved by choosing the witness bunch such that it loads the wakefield to the level before the preceding drive bunch.  In this case, the system becomes periodic, with each subsequent drive bunch re-exciting the wake to its previous amplitude, so that all witness bunches gain the same energy.  If the wakefields are not correctly loaded, each witness bunch will be accelerated to a different energy.  
Although not discussed here, such beams may have applications as probes or be attractive as candidates for compression.

The linear increase of the wakefield amplitude with the number of drive bunches breaks down when the plasma response becomes nonlinear.  As the wakefield amplitude increases, its period increases, such that the drive train no longer resonantly drives the wakefield~\cite{plasma-rosenbluth-saturation}.  Since the plasma wave has a finite radial extent, the wakefield period has a radial dependence, leading to a ``bowing'' of the wakefields~\cite{pwfa-esarey-review}, which further reduces the on-axis accelerating field~\cite{pwfa-lotov-twodimensionalequidistant}.  This dephasing between the drive train and the wakefields it drives results in saturation of the accelerating field.  In this limit, adding more drive bunches will reduce the efficiency of the system.  Even in the absence of nonlinear effects, the longitudinal profile of the drive bunches can reduce the transformer ratio as the wakefields grow~\cite{pwfa-farmer-transformer}.


The key benefit of the novel scheme proposed in this work is that it allows the limitation of saturation to be completely avoided.  Indeed, if the plasma response were fully linear, the same total charge could be accelerated to the same energy as a single witness bunch after the entire drive train.  However, saturation will always play a role in resonantly driven wakefield accelerators if large gradients (close to the cold wavebreaking limit~\cite{plasma-dawson-nonlinear}) are desired.  Even in the limit where the drive train is tailored to compensate the nonlinear lengthening of the wakefield period~\cite{lwfa-umstadter-optimisedlasertrain,lwfa-wettering-modulated}, the wakefield amplitude is still constrained by wavebreaking~\cite{plasma-dawson-nonlinear}.  
The plasma response at high wakefield amplitude therefore limits the energy transfer from the driver to the plasma.  Periodically loading and regenerating the wake avoids this limitation.  Since the spacing between witness bunches in a train is much shorter than the distance between trains of bunches, this allows a higher average power transfer from the drive beam to the plasma, and so to the witness bunches.

In addition to avoiding the limitations of the plasma response to the driver, splitting the witness bunch into a train will reduce transverse beamloading.  This scheme therefore has the potential to accelerate a high average current of positrons. 

To demonstrate the potential of wakefield regeneration within an existing experiment, we consider the case of AWAKE. A proton drive beam self-modulates in plasma, with the resulting train of microbunches resonantly driving a plasma wakefield to high amplitudes. These wakefields are harnessed to accelerate a witness bunch of electrons.  In the forthcoming Run~2c~\cite{pwfa-AWAKE-symmetry}, expected to begin in 2029, two plasma stages will be used with self-modulation of the beam occurring in the first stage.  The witness bunch will then be injected into the second plasma stage, allowing controlled acceleration.  The AWAKE scheme is an ideal candidate for wakefield regeneration, as the proton drive train is already sufficiently long (with an RMS length of several centimetres) for wakefield saturation to occur.  The European Strategy for Particle Physics identified techniques to improve the luminosity of the AWAKE scheme as a key area for future research~\cite{espp2022}.  Altering the length or repetition rate (on the order of 0.1~\si{\hertz}) of the proton driver is an area of active research~\cite{pwfa-farmer-higgs}, but would require significant development of the CERN SPS facility which provides the proton beams used in AWAKE.  The ability to instead accelerate more witness charge per drive beam would increase the achievable luminosity within the timeline of the AWAKE experiment.

Simulations were carried out using the quasistatic particle-in-cell code LCODE~\cite{pic-lotov-lcode,pic-lcode-manual}.  The 400~\si{\giga\electronvolt} proton beam, with a total population of \si{\num{3e11}}, an RMS length of 7.5~\si{\centi\metre}, a radius of 175~\si{\micro\metre} and a normalised emittance of 2.9~\si{\micro\metre}, is propagated through plasma.   The beam is cut at a position 7.5~\si{\centi\metre} (one sigma) ahead of its centre, equivalent to seeding the plasma wakefields with a relativistic ionization front~\cite{pwfa-batsch-seeded}, used to ensure a reproducible bunch train.  The plasma has a density of \si{\num{7e14}~\centi\metre^{-3}}, with a 3\% density step after 1.25~\si{\metre} to avoid the decay of the wakefields after self-modulation~\cite{pwfa-lotov-step}.

\begin{figure}
    \centering
    \includegraphics[width=\linewidth]{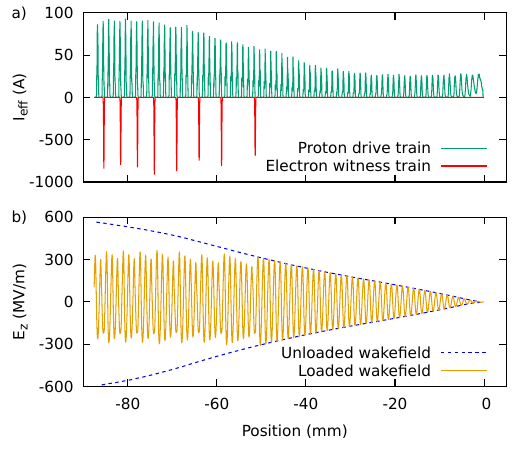}
    \caption{a) The effective current $I_{eff}$ of the proton drive train (green) and electron witness train (red) 10~\si{\metre} into the plasma.  The long proton beam has self-modulated due to its interaction with the plasma, while the train of eight electron bunches has just been injected.  b) The longitudinal wakefield $E_z$ averaged over the 1~\si{\metre} acceleration length (10 - 11~\si{\metre}) for the unloaded (blue dashed line, envelope only) and loaded (yellow line) cases.}
    \label{fig:wakes}
\end{figure}

The proton beam after 10~\si{\metre} propagation
is shown in \ref{fig:wakes}a, in the form of the effective current, 
$I_\mathrm{eff}=\int q\operatorname{K}_0(k_p r)\dif t$, with $q$ the particle charge, $r$ the distance from the axis and $\operatorname{K}_0$ the zeroth-order modified Bessel function of the second kind.  This gives a measure of how strongly each beam slice drives plasma wakefields~\cite{pwfa-katsouleas-beamloading}.  As can be seen, the long drive beam has been modulated into a train of microbunches due to its interaction with the plasma.  The envelope of the unloaded wakefields are shown in Fig.~\ref{fig:wakes}b.  The sub-linear scaling of the wakefield amplitude with the number of bunches shows the influence of the nonlinear plasma response, leading to saturation.

Since the wakefields which drive self-modulation grow along the proton beam, the microbunches are nonuniform.  
The idealised, fully periodic case illustrated in Fig.~\ref{fig:schematic} is therefore not possible.  Nevertheless, monoenergetic acceleration of the witness bunch train is achieved by tailoring the position and charge of the individual witness bunches.

A train of eight witness electron bunches is injected after 10~\si{\metre}.  Each bunch has a Gaussian profile with an RMS length of 60~\si{\micro\metre}, an initial energy of 150~\si{\mega\electronvolt}, and a normalised emittance of 2~\si{\micro\metre}. Witness bunches with a charge of $\sim100$~\si{\pico\coulomb} have been chosen so that the charge density is sufficiently high to drive a blowout, allowing the emittance of the witness bunches to be controlled during acceleration~\cite{pwfa-olsen-inject}.  While the charge density is high, the total charge is sufficiently low that the blowout is localized to the witness location and does not significantly disrupt the wakefields for subsequent bunches~\cite{pwfa-rosenzweig-quasinonlinear}.

The effective current of the witness bunch train at injection is shown in Fig.~\ref{fig:wakes}a, and the corresponding loaded wakefields shown in Fig.~\ref{fig:wakes}b.  Rather than inject into adjacent accelerating buckets of the wakefield, a witness bunch is injected every few plasma periods to keep the witness charge at the desired level 
without overloading the wake.  This also reduces the size of the optimization problem.  The effective current of the driver increases along the length of the witness beam, so later witness bunches are injected more closely together.

\begin{figure}
    \centering
    \includegraphics[width=\linewidth]{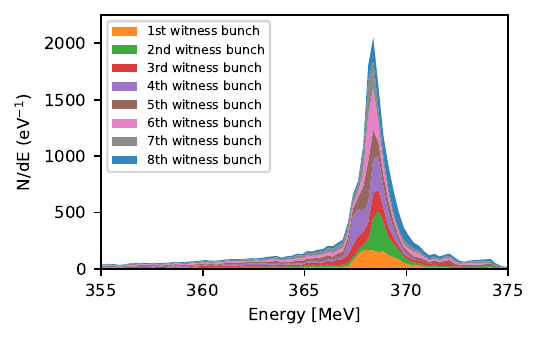}
    \caption{The energy spectrum of the 912~\si{\pico\coulomb} witness train in Fig.~\ref{fig:wakes} after 1~\si{\metre} of acceleration.  The contribution of the eight witness bunches is shown. 56\% of the accelerated charge is within a $\pm0.5$~\% energy range.}
    \label{fig:energy}
\end{figure}

The first witness bunch has a charge of 100~\si{\pico\coulomb}, with the charge of subsequent bunches adjusted such that the loaded wakefields give the same energy gain for all bunches.
The energy gain after 1~\si{\metre} of acceleration, chosen to facilitate a rapid optimization, is shown in Fig.~\ref{fig:energy}.
A total witness charge of 912~\si{\pico\coulomb} is injected and accelerated to 368~\si{\mega\electronvolt}, with 56\% of the accelerated charge within a $\pm0.5$~\% energy range.  The optimal witness train has some dependence on the acceleration length due to the slow evolution of the driver, and so acceleration over longer distances would require the optimization be repeated.  
The density step was chosen to maximize the length of the resulting drive train, and so the acceleration gradient could likely be improved through further optimization of the self-modulation stage.
Injecting later in the drive beam or injecting a lower witness charge would also improve the energy gain.

The benefit of the wakefield-regeneration scheme is immediately apparent from Fig.~\ref{fig:wakes}b.  After the wakefield is loaded by a witness bunch, its amplitude recovers as subsequent drive bunches replenish the wake.  However, the growth rate during this recovery is larger than the growth rate of the unloaded wake.  This is precisely due to the onset of saturation at larger wakefield amplitude.  In the absence of loading, the wakefields gradually dephase with the drive train, and the growth is reduced.  By loading the wakefields, this frequency shift is reduced, and so the drive train remains in phase with the wakefields.  Although not included in these simulations, periodic beamloading will also reduce the impact of ion motion, which in the case of a long drive train is driven by the wakefields themselves and scales with the wakefield amplitude squared~\cite{pwfa-spitsyn-ionmotion,pwfa-turner-ionmotion}.

The unloaded wakefields do continue to grow after the chosen injection point, albeit at a lower rate.  Injecting a single witness bunch at a later position would therefore allow a higher charge to be accelerated.  The unloaded wakefield amplitude, $E_{z0}$, at the position of the eighth witness bunch is 1.9 times that at the position of the first.  The accelerating gradient after beamnloading is $E_z=E_{z0}\left(1-\eta_\mathrm{load}\right)$, where $\eta_\mathrm{load}$ is the fractional beamloading, 29\% for the first witness bunch.  Injecting a single bunch at this later position, then, a beamloading fraction of 63\% could be sustained while delivering the same accelerating field.  Noting that the bunch charge varies as $Q\sim \eta_\mathrm{load} E_{z0}$, it follows that a single witness bunch with a charge of $400$~\si{\pico\coulomb} could be injected and reach the same energy as the witness bunch train simulated here.  The injection of eight witness bunches with a combined charge of 912~\si{\pico\coulomb}, exploiting the enhanced growth rate of the loaded wake, therefore represents a doubling of the charge accelerated in a single shot.  We note that only a fraction of the drive beam is used in these simulations, with the eighth witness bunch approximately at the centre of the drive beam.  Further gains could readily be achieved by adding more witness bunches, exploiting more of the proton beam.

\begin{figure}
    \centering
    \includegraphics[width=\linewidth]{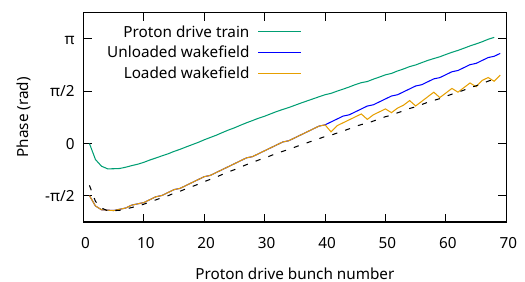}
    \caption{Phase of the proton drive bunch train (green) and the wakefields it excites, after 1~\si{\metre} acceleration.  Both the unloaded (blue) and loaded (yellow) wakefields are shown.  The drive train phase is calculated from the bunch centroids, wakefields as the peaks of the electric field, both taken relative to the first microbunch.  The black dashed line shows a constant offset of $0.8\pi$ from the phase of the drive train.  When the wakefields are loaded, the phase difference between the drive train and wakefields remains near constant.}
    \label{fig:rephase}
\end{figure}

The dephasing between the driver and wakefields is further explored in Fig.~\ref{fig:rephase}, which shows the phase of the microbunch train and the wakefield it drives.
The periodicity of the microbunch train is slightly longer than the resonant plasma frequency due to the growth of the self-modulation instability~\cite{pwfa-pukhov-selfmodulationphase,pwfa-schroeder-growthandphasevelocity}, and the weakly nonlinear plasma response during the self-modulation growth.  The phase of the wakefields follows the same general trend, although the relative phase of the driver and the unloaded wakefields gradually varies along the length of the drive train.  It is this dephasing which leads to the reduction in the wakefield growth.  For the case of the loaded wakefield, the phase difference is kept at a roughly constant level.  This allows the drive train to resonantly excite the wakefields along its entire length, as demonstrated in Fig.~\ref{fig:wakes}b by the enhanced wakefield growth after each witness bunch.


The simulation results show that a train of witness bunches allows more energy to be transferred from the driver to the witness. However, the practical implementation of such a beam may prove challenging. The electron source envisioned for AWAKE Run~2c incorporates an S-band RF photo injector~\cite{pwfa-pepitone-injector}, which would prevent the injection of witness bunches with a spacing of a few millimetres.  One solution, albeit a costly one, would be to incorporate multiple such electron guns into the experimental setup.  Alternative injection schemes, such as laser-foil injection~\cite{pwfa-khudiakov-laserassisted,pwfa-wilson-laserassisted}, or laser-plasma injection~\cite{pwfa-minenna-earli}, would potentially allow the creation of a suitable train of witness bunches by using a train of laser pulses.  Within the constraints of the currently forseen Run~2c budget, it may be possible to accelerate two witness bunches within the same RF bucket, although this would require further study.

The optimization itself is relatively straightforward, as the charge and position of each subsequent witness bunch is essentially tuned independently until they reach the energy of the preceding bunches.  Since the optimization must be redone for each witness bunch, but the optimization itself will be similar, this would be a good candidate for machine learning techniques.

As discussed above, the requirement for a nonuniform witness train is a direct consequence of the drive beam in the AWAKE scheme.  The use of driver made up of uniform bunches would allow the use of a uniform witness train, significantly easing the constraints on tuning.  For the SPS proton beam, such a drive train could potentially be achieved using a dielectric bunching structure~\cite{pwfa-petrenko-modulate}.  Laser schemes such as the plasma-modulated plasma accelerator~\cite{lwfa-jakobsson-modulated} or electro-optic frequency combs~\cite{laser-ye-comb,laser-yang-cavitycomb} could also provide a suitable driver.  Periodic electron drivers could be generated using a train of laser pulses on a photocathode~\cite{pwfa-andonian-argonnetrain}, or by modulating an electron beam using a mask~\cite{pwfa-muggli-mask}, a terahertz cavity~\cite{pwfa-hibberd-terahertz}, or inverse-FEL bunching~\cite{pwfa-manwani-ifel}.

In summary, we propose a new scheme of plasma wakefield acceleration, where a train of witness microbunches are accelerated in the wake driven by a train of drive bunches.  Key to this concept is that periodically loading the wakefields prevents the dephasing which occurs in resonantly driven wakefield accelerators as the plasma response becomes nonlinear.  The energy transfer from the driver to the plasma can therefore be significantly increased, allowing a larger total witness charge to be accelerated per drive train, increasing the luminosity.  Proof-of-concept simulations for the AWAKE experiment demonstrate the acceleration of a 912~\si{\pico\coulomb} train of witness electron bunches, approximately twice that which could be accelerated in a single bunch.  Further gains are readily available through the addition of more witness bunches.  This method is generally applicable to resonantly driven wakefield accelerators, and so could equally be applied to other schemes such as the plasma-modulated plasma accelerator~\cite{lwfa-jakobsson-modulated}.  The scheme also permits a high average witness current while keeping the peak witness current low, desirable for plasma-based positron acceleration.

\begin{acknowledgments}The authors would like to thank Eduardo Granados, who first proposed this topic, and Konstantin Lotov and Patric Muggli for fruitful discussions.  JPF gratefully acknowledges the Gauss Centre for Supercomputing e.V. (www.gauss-centre.eu) for funding this project by providing computing time through the John von Neumann Institute for Computing (NIC) on the GCS Supercomputer JUWELS at Jülich Supercomputing Centre (JSC).
\end{acknowledgments}

\bibliography{bib}

\end{document}